# On-surface dehydrogenative lateral homo-coupling and aromatization of n-octane on Pt(111)

*D. Arribas[1,2]*, E. Tosi[1], V. Villalobos-Vilda[1], B. Cirera[1], I. Palacio[1], A. Sáez-Coronado[1], P. Lacovig[3], A. Baraldi[3,4], L. Bignardi[4], S. Lizzit[3], C. Sánchez-Sánchez[1,7], A. Gutiérrez[5,6], J. A. Martín-Gago[1,7], M. Garnica [6,7,8], J. I. Martínez[1,7], P. L. de Andres[1], P. Merino[1,7]**

1) Instituto de Ciencia de Materiales de Madrid, c/ Sor Juana Inés de la Cruz, 3, 28049 Madrid, Spain

2) Max-Planck-Institut für Festkörperforschung, Heisenbergstraße 1, Stuttgart, 70569, Germany

3) Elettra-Sincrotrone Trieste S.C.p.A, S.S. 14 km 163.5, Trieste, Italy

4) Physics Department, University of Trieste, Via Valerio 2, 34127 Trieste, Italy

5) Universidad Autónoma de Madrid, c/ Francisco Tomás y Valiente, 7, 28049 Madrid, Spain

6) Instituto Nicolás Cabrera, Cantoblanco, Calle Francisco Tomás y Valiente, 7 28049 Madrid, Spain

7) Unidad de Nanomateriales Avanzados, IMDEA Nanoscience, Unidad asociada al CSIC por el ICMM, Madrid 28049, Spain

8) Instituto Madrileño de Estudios Avanzados en Nanociencia (IMDEA-Nanociencia), Madrid, Cantoblanco 28049, Spain

* Corresponding authors: D. Arribas: d.arribas@fkf.mpg.de, P. Merino: pablo.merino@csic.es

ABSTRACT: Aliphatic hydrocarbons, such as normal alkanes (*n*-$C_nH_{2n+2}$), constitute a naturally abundant source of carbon atoms. Of special interest is the formation of cyclic and aromatic products from aliphatic reactants. Combining scanning tunneling microscopy and *ab-initio* calculations, we investigate the thermal-induced aromatization of linear *n*-octane (*n*-$C_8H_{18}$) molecules on the catalytic Pt(111) surface and the reactions of intermolecular homo-coupling between them at temperatures above 600 K. The cycloaromatization of individual *n*-octane molecules requires bending the linear adsorbates prior to their dehydrogenation and the formation of an intramolecular C−C bond, yielding adsorbed benzene rings. In addition, the Pt(111) surface catalyzes a homo-coupling reaction by initiating the formation of a C−C bond between the dehydrogenated methyl ends of the chemisorbed *n*-octane molecules and then propagating along the carbon backbone in a zipper-like fashion. Our findings provide molecular-level insight into the heterogeneous catalytic processes underlying the generation of aromatic products and stable on-surface polycyclic species.

## 1-INTRODUCTION:

The family of alkanes include those hydrocarbons consisting of $sp^3$-hybridised carbon atoms fully saturated with hydrogen. Although they constitute an ideal carbon source for more complex organic compounds, the intrinsic lack of reactivity of the $C(sp^3)$−H bonds hinders their direct use as precursors without prior functionalization.[1–5] In this regard, on-surface chemistry provides a powerful set of strategies to synthesize organic nanostructures by bottom-up approaches from a wide range of precursors.[6,7] For alkanes, the linear C−C coupling of unbranched *n*-alkanes has been achieved on Au(110) and gold-covered Ag(110) surfaces.[8–10] This results in the formation of long alkane chains, thanks to the one dimensional confinement of the molecules between the parallel surface ridges of these substrates.[10,11] In contrast, on the more reactive Cu(110) surface, *n*-alkanes first undergo a thermal-induced cascade dehydrogenation reaction, leading to the formation of polyolefin chains that subsequently couple at their ends to form longer chains.[12,13]

Of great interest is the synthesis of (poly)cyclic hydrocarbon compounds, which are the base for a considerable part of recent works in on-surface chemistry.[14–18] (Poly)cyclic and (poly)aromatic molecules exhibit appealing properties, finding applications in fields such as molecular electronics,[19,20] carbon-based molecular spintronic,[21–23] metal-organic frameworks[24] or organic light emitters[25] among others.[26] However, the controlled on-surface synthesis of cyclic and aromatic species from linear aliphatic precursors requires a two-dimensional growth that is hindered by the ridges typically present on (110) surfaces.[11] In addition, the high temperatures required to activate these reactions promote the competition with uncontrolled side reactions, such as the fragmentation of the reactants, especially in the case of longer alkanes.[27] From the fundamental point of view, the coexistence of multiple intricate reaction pathways renders the characterization of the on-surface aromatization process particularly challenging.

In principle, there exist two possible routes leading to the formation of cyclic or aromatic products from linear *n*-alkanes. The first one involves the intramolecular rearrangement of individual reactant molecules (catalytic reforming).[27] Depending on the number of carbon atoms in the reactant, cyclic or aromatic products of different sizes and structures can be obtained. This is the route typically followed in the catalytic cycloaromatization of hydrocarbons on supported catalysts.[28–30] The second route implicates the chemical coupling of several reactant molecules that coalesce after adsorption on the surface. For example, several intermolecular C−C bonds can be formed between two partially dehydrogenated molecules arranged side by side. This process can involve several molecules, ultimately yielding large polyaromatic species after complete dehydrogenation of the product. As these dehydrogenative processes are highly endothermic in general, the choice of an appropriate substrate plays a crucial role as it must be catalytically active but also stabilize the precursor to prevent its desorption at the reaction temperature.

Herein, we investigate the two aforementioned routes using linear *n*-octane ($n$-$C_8H_{18}$) molecules adsorbed on a catalytic Pt(111) surface as a model system. Platinum is a common catalyst for dehydrogenation of alkanes as it exhibits a higher selectivity towards the dehydrogenative processes than towards C−C bond cleavage.[31,32] Deposition and annealing at mild temperatures of $n$-$C_8H_{18}$ on Pt(111) results in the chemisorption and partial dehydrogenation of the adsorbates, yielding unsaturated *n*-octenyl groups as revealed by synchrotron-radiation X-ray photoelectron spectroscopy (XPS) and Scanning Tunneling Microscopy (STM).[33] In this work, we focus on the reactions involving these unsaturated intermediates at temperatures between 600 K and 1000 K.

According to our submolecularly resolved STM images, combined with first-principles density functional theory (DFT)-based calculations and STM-imaging simulations, the individual linear precursors can undergo a dehydrogenative cyclisation, ultimately leading to the formation of planar hexagonal rings of six carbon atoms (see **Scheme 1**). On the other hand, when two precursors find each other prior to individual cyclisation, the high thermal energy supplied leads to intermolecular side-reactions in which linear adsorbates arrange parallelly and form multiple C−C bonds in a sequential homo-coupling process, yielding polycyclic products of the anthracene family. Subsequent diffusion and aggregation of the adsorbates lead to an increase in the size of the clusters, eventually resulting in the formation of extended nanographenes.[34]

**Scheme 1.** *n*-Octane aromatization reaction and homo-coupling side reaction on Pt(111)

## 2-RESULTS AND DISCUSSION:

To conduct the experiments presented below, *n*-octane was deposited on a pristine Pt(111) surface under ultra-high vacuum conditions by physical evaporation. At room temperature *n*-octane readily evaporates from an external evaporator connected to the vacuum chamber with a leak valve allowing control of the dose of molecules being deposited (more details in the **Experimental and computational methods** below). If the deposition takes place at room temperature, *n*-octane physisorbs on the surface without chemical modification (see **Fig. 1b**), appearing as a linear feature with a length of approximately 10 Å in STM images (**Fig. 1a**). Annealing of the sample above 330 K produces a highly regio-selective cleavage of a C−H bond at one of the methyl ends of the molecules (see **Fig. S1** in the Supporting information),[33] followed by further dehydrogenation of its carbon backbone. This yields unsaturated *n*-octenyl groups chemisorbed on the surface, denoted as **1** in **Fig. 1 g**.[35,36]

## 2.1- Cyclization and aromatization of individual molecules

After annealing at high temperature (at 890 K) we find rounded features whose size is comparable to that of an individual *n*-octane molecule after cyclisation (see **Fig 1b** and **S3** in the Supporting information). Atomic-resolution STM images show features exhibiting a diameter of approximately 2.5 Å and an apparent height of 0.5 Å, displaying a three-fold intramolecular symmetry and a depression in their center. These dimensions are comparable to the interatomic distance of the Pt(111) substrate, which suggests that each feature originates from a single *n*-octane molecule. Its adsorption position, deduced from the atomically resolved substrate lattice in **Fig. 1c**, is on hollow, with three brighter vertices pointing along the second-neighbor Pt directions (see also **Fig. S3c**).

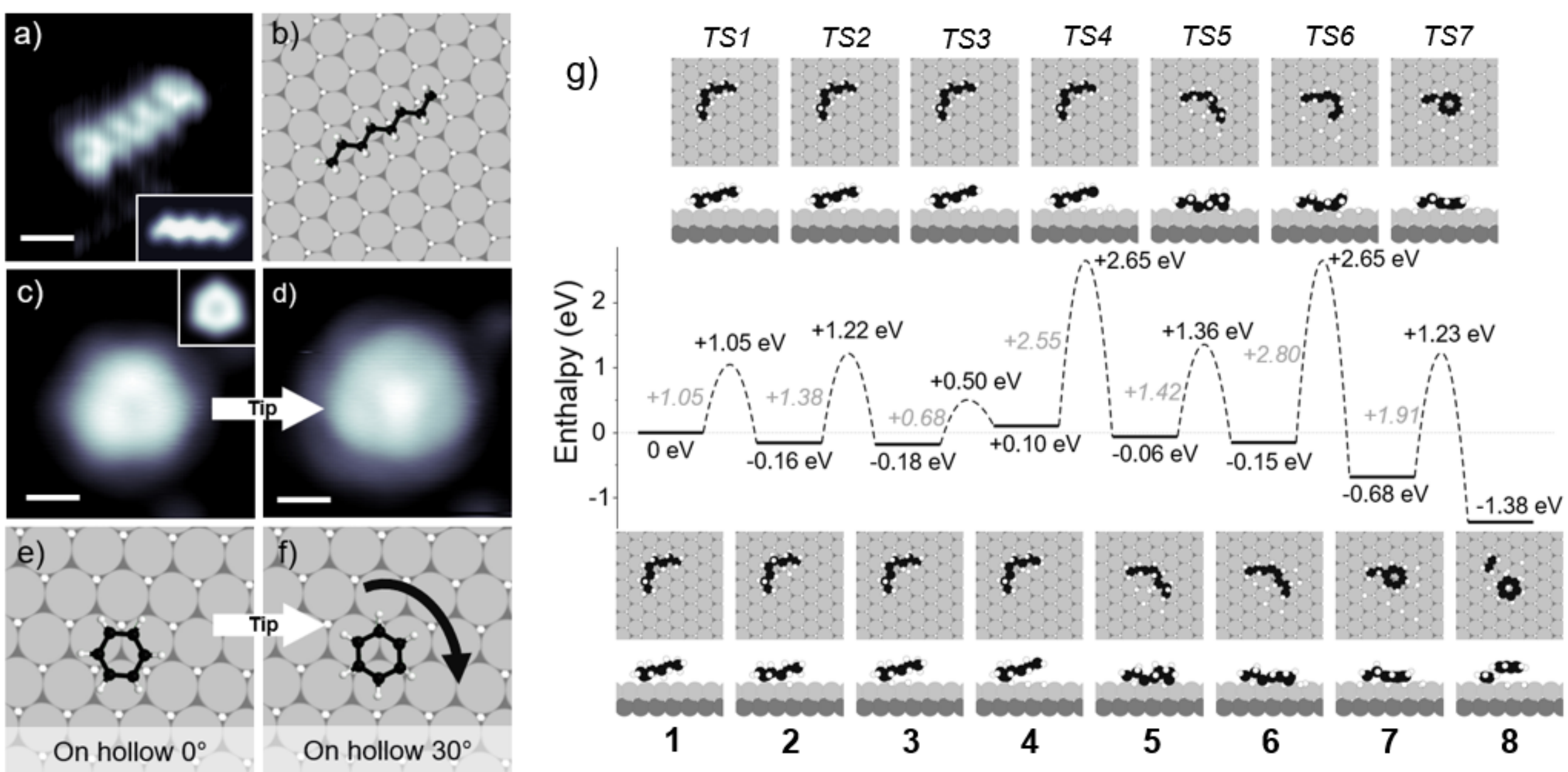

**Figure 1:** a) Submolecular-resolution STM image (2 nA, -20 mV, 25 Å × 25 Å, scale bar corresponds to 5 Å, acquired at 77 K) of an unreacted *n*-octane molecule after on Pt(111) at 390 K. The inset contains a simulated STM image of an unreacted *n*-octane on Pt(111). b) Schematic representation of the unreacted molecule in panel a). c) High resolution STM of benzene ring obtained after thermal treatment at 900 K (0.1 nA, -50 mV, 16 Å × 16 Å, scale bar corresponds to 3 Å, acquired at 77 K). d) Previous benzene ring after manipulation with the tip, resulting in a change of adsorption configuration (0.05 nA, -50 mV, 16 Å × 16 Å, scale bar corresponds to 3 Å, acquired at 77 K).[38,39] e) and f) Schematic representation of the adsorption configurations in c) and d). g) Enthalpic path of the aromatization reaction together with the atomic representations of the reaction steps and intermediates.

Comparison of our STM images with simulated Tersoff-Hamann STM images for the DFT structural optimization of a benzene molecule adsorbed at a hollow site shows good agreement (inset in **Fig. 1c**). Therefore, we assign these structures to benzene molecules derived from *n*-octane dehydrogenation adsorbed on the Pt(111) surface.[37–39] This assignment is also consistent with previous studies on benzene adsorption on Pt(111), which reported that benzene is stable at room temperature on Pt(111) and its shape in STM images depends on the exact adsorption site,[38] adopting a triangular shape when it adsorbs on-hollow on Pt(111),[39] which is one of the most stable configurations.[40] The absence of protrusions around the benzene ring implies that there are no methyl or ethyl groups attached to it, as it would be the case, for example, in *para*-xylene deposited on Pt(111).[41] As *n*-octane has eight carbon atoms, two more than benzene, this implies that two carbon atoms have been cleaved during the cyclization process.

To gain further insight into the reaction pathway resulting in the formation of benzene rings from linear *n*-octane molecules, we complement our STM observations with DFT-based calculations of the reaction intermediates, transition states (TS) and their enthalpy barriers (**Fig. 1**g). According to our STM observations, *n*-octane initially physisorbs on the Pt(111) surface in an *all-trans* configuration. The first C−H bond cleavage takes place at one methyl end yielding a chemisorbed *n*-octyl group covalently bonded to the Pt substrate. Subsequent dehydrogenation in the carbon backbone induces the formation of a double bond between the 3rd and 4th carbon atoms, yielding an *n*-octenyl group in a slightly endothermic process (+0.1 eV).[33] To proceed with the reaction, the linear adsorbate must bend to adopt a more compact configuration (**1**). This isomerization is promoted by the higher rotational freedom of the remaining $sp^3$ bonds in the precursor and leads to sequential dehydrogenations (**2-4**). The formation of a C−C bond between the terminal C atom and the third C atom in the chain (**TS6**) constitutes a limiting step of the reaction, associated with an enthalpy barrier of +2.8 eV. In the experiment, the hydrogen released in the process diffuses or desorbs owing to the high thermal energy provided at the reaction temperatures. Therefore, after the formation of the phenyl moiety, the ring is stable and cleavage of the newly formed C−C bond by rehydrogenation is unlikely. These dehydrogenative reactions of *n*-octane continues with the dehydrogenation of the hexagonal ring, which leads to its aromatization and the formation of ethylbenzene (**7**) in an overall exothermic process (-0.7 eV). Finally, to recover the experimentally observed benzene adsorbate (**8**), the ethyl group must be cleaved from the aromatic ring (**TS7**). This step proceeds through the intercalation of a H atom between the C atoms connecting the aromatic ring with the ethyl group (intermediate **2d**) to form $C_2H_4$, for which a reaction barrier of 1.9 eV must be overcome. The 2-carbon fragment released in this step most likely diffuses away from the benzene or desorbs from the substrate at the high temperatures required for the Pt-catalyzed reaction.

Overall, the thermal aromatization of *n*-octane on Pt(111) is an exothermic process with a total enthalpy variation of -1.4 eV and a rate limiting step associated with an activation barrier that we estimate as +2.8 eV. In general, the reactions proposed in **Fig. 1** are feasible only thanks to the high reaction temperature in the real system. However, a rough estimation of the reaction rate using Arrhenius' law with an activation energy barrier of +2.8 eV and a preexponential factor of $10^{13}$ $s^{-1}$ yields a reaction rate of $1.9 \cdot 10^{-3}$ $s^{-1}$. This slow rate indicates that other competing reactions being more favorable and explains the small number of unimolecular cyclization products observed in our STM images compared with adsorbate aggregation and intermolecular reactions (see **Fig. S3** in the Supporting Information).

### *2.2- Lateral homo-coupling of* n-*alkanes*

High temperatures favor the diffusion of the deposited adsorbates across the Pt(111) surface, promoting aggregation and intermolecular reactions (according to our experimental observations, *n*-octane-derived adsorbates only aggregate at temperatures above 650 K, compare **Sections 1** and **2** from the Supporting Information). Since the final product of the dehydrogenation are nanographenes containing a larger number of C atoms than the single molecules (see **Section 4** in the Supporting Information), there must be a set of chemical reactions leading to the coupling of the carbon atoms coming from different alkane chains. To avoid the complexity associated with the characterization of the large multimolecular clusters appearing at temperatures above 650 K (see **Fig. S2** and **S3** in the Supporting Information), we focus on samples annealed at 600 K in which the aggregation is still inefficient. In this temperature range we observe the formation of molecular dimers (**Fig. 2c** and **d**), which take the aspect of an elongated lobed structure with a minimum nodal line that traverses its longitudinal axis (see **Section 3** of the Supporting Information).

The good agreement between experimental and DFT-based simulated STM images (see **Figure 2k** and **l**) supports the assignment of the observed product in **Figure 2c** and **d** as an adsorbed anthracene molecule or to a derivative of this molecule (the configuration in **Fig. 2l** corresponds to anthracene, for the STM image simulations of other similar candidate molecules see **Fig. S5**), in which the two original *n*-octane molecules have established four intermolecular C-C bonds. In addition, tip-induced manipulations of this type of dimer without dissociation of the individual chains points to the existence of strong covalent interaction between them. Due to its radical character, the activated C-H bond at the methyl end of the chemisorbed *n*-octyl and/or *n*-octenyl groups constitutes the part of the individual chains most prone to start the reaction. Thanks to the thermally induced on-surface diffusion, the approximation of two suitably oriented *n*-octyl radicals can trigger the formation of a covalent C-C intermolecular bond between their first carbon atoms of each molecular chain, similar to the end-to-end bonding that takes place in the linear polymerization of alkanes on Au(110).[11]

The observation of molecular hydrocarbon aggregates in our STM images, with shapes and sizes consistent with the lateral fusion of a pair of molecules, suggests that this initial intermolecular bond is the first step of a series of dehydrogenative homo-coupling reactions resulting in the apparition of several C−C bonds connecting the carbon backbones of adjacent adsorbates in a homo-coupling process. In fact, tip-induced manipulation of individual *n*-octyl groups chemisorbed on Pt(111) shows that it is possible to approximate the linear adsorbates, which under certain scanning parameters appear as four lobed elongated features (see also **Fig. S4**), increasing the interaction between their activated methyl ends (see **Fig. 1a**, **b**, **e**, **f**). Upon annealing the samples at 650 K or above, the high thermal energy available allows for the encounter of molecular pairs and triggers the subsequent homo-coupling reaction.

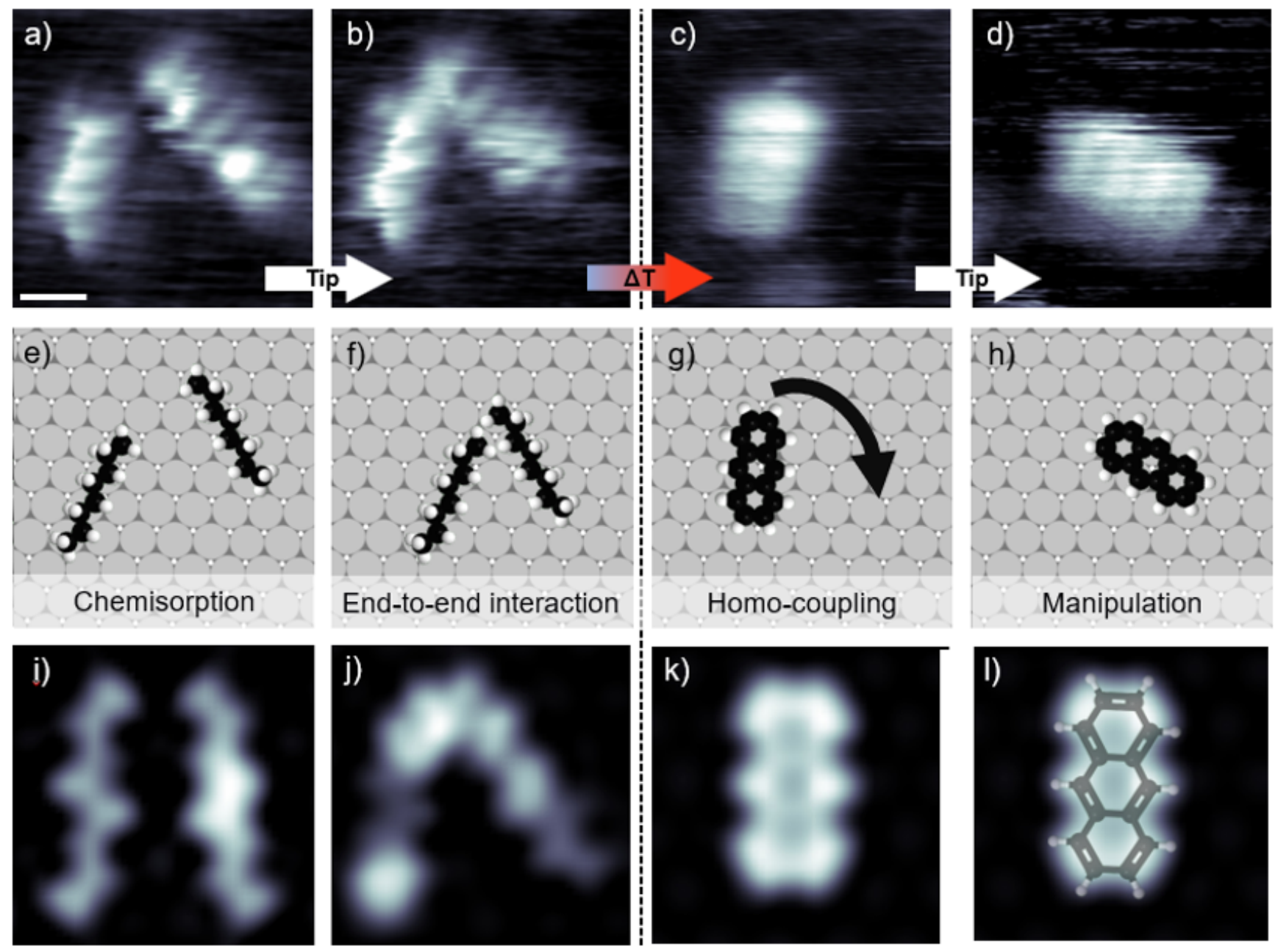

**Figure 2.** a) STM image of two *n*-octyl groups chemisorbed to close positions on the surface (0.78 nA, -25 mV, 22 Å × 22 Å, room temperature, the scale bar corresponds to 5 Å). The sample was annealed at 600 K prior to room temperature STM measurements. b) Due to the interaction with the tip during image acquisition, both molecules move closer and interact with each other through their extremes (0.78 nA, -25 mV, 22 Å × 22 Å, room temperature). c) STM image from a different sample flash-annealed at 890 K in which the thermal energy was enough to form C-C bonds between two molecules (0.32 nA, -84 mV, 22 Å × 22 Å, room temperature). d) Displacement without dissociation of the dimer induced by interaction with the tip (0.10 nA, -10 mV, 22 Å × 22 Å, room temperature), highlighting the strong interaction between both chains. e), f), g) and h) are the representations of the atomic configurations in the previous panels. In the third row, simulated STM images at -0.1 V of i) two fully unsaturated *n*-tetraoctenyl groups, j) two *n*-tetraoctenyl groups covalently bonded at their methyl ends, k) anthracene, obtained after completion of the homo-coupling process and cleavage of a two-carbon fragment,. l) Simulated STM image in k) with the atomic configuration of the adsorbate superimposed.

To gain further insight into the atomistic mechanisms involved in the homo-coupling reaction (**Figure 3**), we have conducted DFT-based simulations. We take as an initial state two *n*-octa-1,3,5,7-tetraenyl groups (denoted as **1** in **Fig. 3e**) chemisorbed in close proximity on Pt(111), which correspond to fully unsaturated eight-$sp^2$-carbon chains. In our simulations, the reaction initiates by sequential dehydrogenation (species **2**, **3**) with activation barriers of ≈1 eV comparable to those calculated for the isolated molecule cyclization pathway. After removal of six H atoms, the partially dehydrogenated fragments (**4**) approach each other and initiate the formation of two closed carbon rings (**5**). Along this pathway, we identify a transition state (**TS4**) with an energy barrier of +2.23 eV. Further dehydrogenation and structural rearrangement promote closure of the third fused ring (**6**), thereby generating the carbon backbone of anthracene. This ring-closure step introduces a second kinetically limiting barrier (**TS5**) of ≈2.8 eV. In contrast to the preceding barrier, which could in principle be lowered through a multistep sequence, this latter barrier appears to arise from the concerted structural rearrangement required to complete the third ring. Finally, detachment of a $C_2H_4$ fragment (**8**) completes the transformation. In this process, the interaction with the Pt substrate plays a crucial role in stabilizing the intermediates (in agreement with synchrotron radiation X-ray photoelectron spectroscopy measurements in **Section 5** of the Supporting Information) and favoring the dehydrogenation of the C atoms involved in the reaction. Remarkably, temperature-programmed desorption measurements reveal a desorption signal at *m/z* = 28 around 900 K, consistent with the release of $C_2H_4$ byproducts generated during benzene cyclization and anthracene-forming homo-coupling pathways (Fig. 3f). In terms of enthalpy, the reaction described in **Figure 3** is exothermic, yielding an overall energy gain of ≈−1.8 eV relative to the initial state with a rate-limiting step associated with an estimated energy barrier of +2.8 eV, which can be overcome under the high annealing temperatures in our experiments. This surface-catalyzed homo-coupling reaction between aliphatic precursors yields polycyclic compounds containing hexagonal rings with the potential of becoming larger polyaromatic compounds after further coupling with other adsorbates.

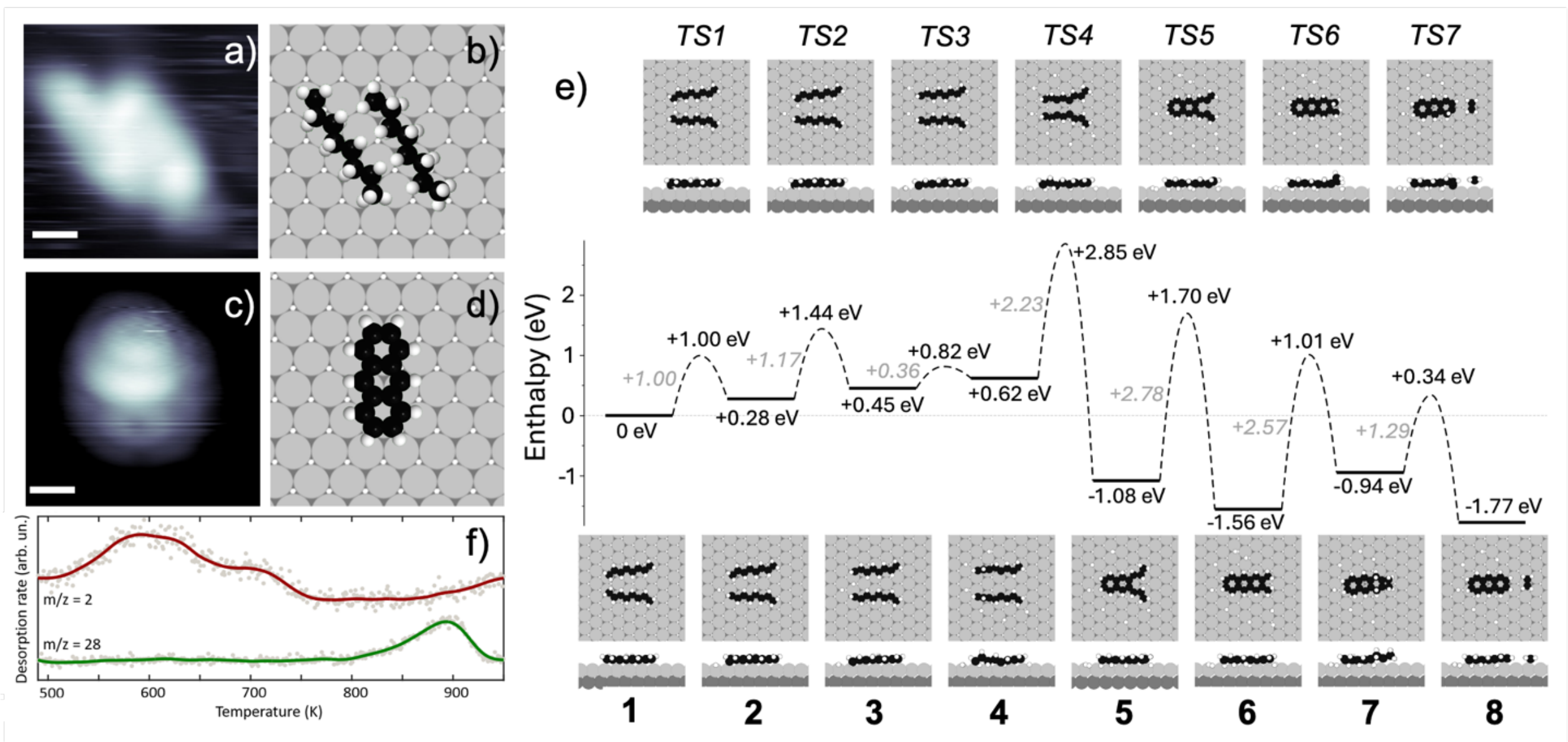

**Figure 3.** a), b) High resolution STM image and schematic representation of two unreacted *n*-octyl groups after deposition at 390 K (1 nA, 1000 mV, 22 Å × 22 Å, scale bar corresponds to 5 Å, acquired at 77 K) interacting laterally prior to initiation of the homo-coupling reaction. c), d) High resolution STM image and schematic representation of an athracene product obtained after thermal treatment at 850 K (0.02 nA, -250 mV, 22 Å × 22 Å, scale bar corresponds to 5 Å, acquired at 77 K). e) Enthalpic path of the aromatization reaction together with the atomic representations of the reaction steps and intermediates. f) Temperature-programmed desorption measurement of molecular hydrogen (m/z =2) and $C_2H_4$ (m/z = 28) reaction byproducts at temperatures 500 - 1000 K,

In the case of aggregates involving more than two molecules (such as those in **Figure S7 a** obtained at higher temperatures), the homo-coupling mechanism may contribute to the formation of nanographene patches (see **Fig. S7 c** and **Fig. S8** in the Supporting Information) since the size of the molecular aggregates observed in our STM images after annealing above 600 K clearly exceeds that expected for the final product predicted for the lateral coupling of only two aliphatic chains (compare **Figures 2 c**, **d**, and **Fig. S7**). However, the parallel arrangement observed in images with submolecular resolution, and the formation of large graphene flakes after annealing at temperatures above 1000 K (see **Supp. Info.**), suggests that the lateral coupling of more molecules might proceed in a similar zipper-like mechanism as the one described in **Figure 3**. In this case, the dehydrogenated molecules would establish intermolecular bonds with other previously formed polycyclic adsorbates in a sequential molecule-by-molecule fashion.

## 3- CONCLUSIONS

In conclusion, our results reveal how Pt(111) catalyzes the transformation of linear alkanes into cyclic, aromatic, and polycyclic carbon structures through a sequence of C−H activation, molecular rearrangement, and C−C bond-forming steps. By combining submolecular STM imaging with DFT-based calculations, we identify the key intermediates and kinetic bottlenecks governing both unimolecular cycloaromatization and intermolecular dehydrogenative homo-coupling. These findings provide an atomic-scale picture of catalytic reforming processes, highlighting the decisive role of surface-stabilized unsaturated intermediates, adsorbate diffusion, and sequential ring closure in controlling product selectivity. More broadly, this work establishes that on-surface chemistry under well-defined conditions constitutes a powerful approach to disentangle complex hydrocarbon reforming pathways, offering perspectives for the rational design of catalysts capable of steering aliphatic precursors toward selected aromatic products.

## 4- EXPERIMENTAL AND COMPUTATIONAL METHODS:

### 4.1- Experimental details

We conduct our experiments under ultra-high vacuum conditions, in systems with a base pressure below $2\cdot10^{-10}$ mbar. To clean our Pt(111) single-crystals we perform several cycles of $Ar^{+}$ cycles at 1.5 kV followed by temperature flashes at 1100 K, the first of them under an oxidative $O_2$ atmosphere to remove carbon contaminants from the surface. The surface was considered clean once pristine atomically flat terraces delimited by sharps steps are observed by STM or when the characteristic level shift in the Pt-4$f_{7/2}$ XPS spectrum is observed.[42] We deposit the *n*-octane molecules (Sigma-Aldrich, purity >99 %) without further purification by thermal evaporation from an external quartz evaporator at room temperature. To control the dose, we employ a leak valve and keeping the deposition pressure between $0.4 \cdot 10^{-8}$ and $4 \cdot 10^{-8}$ mbar for typical deposition times of 100 seconds, which are enough to saturate the surface. We keep the sample at room temperature or at 77 K (in the case of the samples used for synchrotron radiation XPS experiments) during the deposition. The temperature was measured by means of a thermocouple welded to the back of the sample or with a pyrometer (emissivity: $\varepsilon = 0.1$). We perform our STM measurements at room temperature and low temperature (77 K) in two independent UHV system after *n*-octane deposition on the Pt(111) surface and subsequent annealing to the desired temperature. We use a chemically etched tungsten tip conditioned under UHV by standard procedures. STM images are acquired in constant current mode, and both topographic and current error images are analyzed. To acquire and post-process the images we employ the WSxM sorftware.[43]

The XPS experiments presented in the Supporting Information were conducted at the SuperESCA beamline of the Elettra Synchrotron facility (Italy). We keep the sample at 77 K and measured at normal emission for high-resolution XPS spectra acquisition. In these experiments, a photon energy of 400 eV maximizes the surface sensitivity and the photoionization cross-sections for the C-1s core levels. No beam-induced damage to the samples was detected during spectra acquisition. For spectra deconvolution we subtract a Shirley background and fit each component to a pseudo-Voigt function (a linear combination of a Gaussian and a Lorentzian profiles). Fitting parameters include the binding-energy, the intensity (area), the full width at half maximum (typically around 400 meV), and the Gaussian-Lorentzian ratio. We take special care to obtain consistent values for these parameters in all the spectra. In the Temperature-Programmed Desorption (TPD) experiment described in the Supporting Information, desorbing molecules were detected by a quadrupole mass spectrometer (QMS) whose ionization volume is enclosed within a self-built Feulner cup,[44] featuring an opening comparable in size to the crystal surface. During the temperature ramp (2 K $s^{-1}$), the sample was positioned in front of this opening. This configuration effectively minimizes contributions from residual species in the vacuum chamber and ensures an excellent signal-to-noise ratio.

### 4.2- Computational methods

We have carried out a large battery of Density Functional Theory (DFT)-based calculations and STM-imaging simulations for all the systems involved in the present investigation. Gibbs free energies of gas-phase molecules were calculated using the Gaussian16 suite[45] *via ab-initio* density functional theory (DFT), employing the B3LYP hybrid exchange-correlation functional[46,47] and the cc-pVTZ basis set.[48] Periodic systems were modeled using DFT with soft pseudopotentials[49] and a plane-wave basis set with a cutoff energy of 410 eV, as implemented in the CASTEP code.[50] The electronic exchange-correlation has been accounted for by the GGA-PBE functional.[51] Dispersive forces have been included in the calculations by the Tkatchenko-Scheffler van der Waals correction scheme.[52] Brillouin zones in all the interfacial model systems have been sampled by optimal Monkhorst-Pack [4×4×1] k-point grids.[53] The Pt(111) substrates were modeled as infinite 2D periodic slabs with four physical layers, keeping fixed the two bottommost ones during the geometrical optimizations. Systems in neighboring cells along the perpendicular-to-the-surface direction were separated by at least a 20 Å-thick vacuum region to avoid the interaction between two adjacent slabs. Transition States (TS) were determined using the full quadratic synchronous transit method (QST) (Govind et al., 2003) as implemented in the CASTEP package (Clark et al., 2005). The electronic energy was converged to 10−5 eV using soft pseudopotentials and an energy cutoff of 270 eV. For TS determination, structural optimizations were performed until the maximum residual force on each atom was below 0.05 eV/Å. Brillouin zone sampling was restricted to the Γ point. The calculated diffusion barrier for atomic hydrogen on Pt(111) is approx10 meV, indicating essentially barrierless surface mobility. Consequently, H atoms transferred from the molecule to the surface are expected to diffuse rapidly across the substrate. Unless otherwise stated, re-adsorption or re-association processes of atomic hydrogen onto the molecule were neglected due to the rapid diffusion of these species. In addition, molecular hydrogen dissociation on Pt(111) was found to proceed essentially without an activation barrier, yielding chemisorbed atomic hydrogen with a net energy gain of ≈ 1 eV. The Tersoff-Haman formalism was used for simulation of the STM images.[54]

ASSOCIATED CONTENT

**Supporting Information**.

Further STM characterization of the *n*-octane depositions on Pt(111) after annealing at 600 K (Section 1) and 650 K (Section 2). Further details on the experimental and simulated STM images of the dimer in Figure 2 and dimer dehydrogenation barrier (Section 3). STM characterization of nanographenes obtained after annealing at 1050 K (Section 4). Complementary characterization by synchrotron radiation X ray photoelectron spectroscopy and thermal program desorption of the samples at different temperatures. (DOCX)

## AUTHOR INFORMATION

**Corresponding Author**

D. Arribas: d.arribas@fkf.mpg.de, P. Merino: pablo.merino@csic.es

**Author Contributions**

The manuscript was written through contributions of all authors. All authors have given approval to the final version of the manuscript.

## ACKNOWLEDGMENT

This research was financially supported by projects from Spanish MCIN (Grants PID2024-157655NB-I00, PID2023-149077OB-C31, TED2021-12941A-I00 and PLEC2021-007906 funded by MCIN/AEI/10.13039/501100011033 and by the "European Union NextGenerationEU/PRTR"), and from Comunidad de Madrid (Grant TEC-2024/TEC-459). IMDEA Nanociencia and ICMM-CSIC acknowledge financial support from the Spanish Ministry of Science and Innovation 'Severo Ochoa' Grant CEX2020-001039-S and Grant CEX2024-001445-S Programme for Centres of Excellence in R&D, respectively. M.G. has received financial support through the "Ramón y Cajal" Fellowship program (RYC2020-029317-I), "Ayudas para Incentivar la Consolidación Investigadora" (CNS2022-135175) and the (MAD2D-CM)-MRR MATERIALES AVANZADOS-IMDEA-NC project. D. Arribas acknowledges the Spanish Ministry of Universities, FPU-2019 predoctoral grant (FPU19/04556).

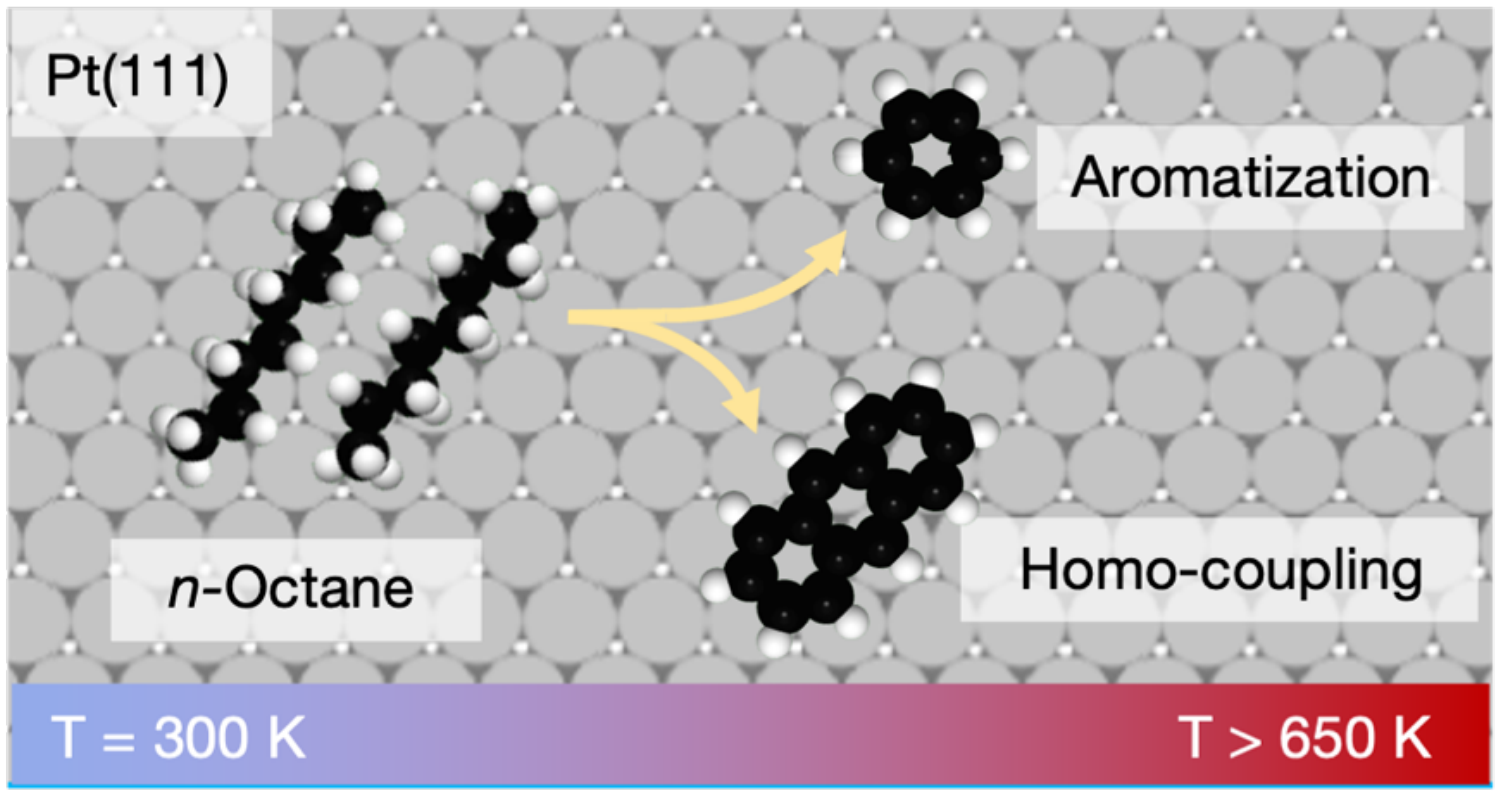

Table of Contents